\documentclass[fleqn,usenatbib]{mnras}
\usepackage{newtxtext,newtxmath}
\usepackage[T1]{fontenc}
\usepackage{ae,aecompl}

\usepackage{color, soul, ulem,graphicx,commath}

\newcommand{\msano}{{\rm M}_\odot ~{\rm yr}^{-1}}
\newcommand{\mdot}{\dot{M}}
\newcommand{\gj}{GJ\,436}
\newcommand{\ly}{Ly-$\alpha$}
\defcitealias{2016A&A...591A.121B}{B16}

\newcommand{\e}[1]{\times 10^{#1}}

\title{Exoplanets as probes of the winds of host stars: the case of the M dwarf GJ 436}
\author[A.~A.~Vidotto \& V.~Bourrier]{A.~A.~Vidotto$^{1}$\thanks{E-mail: Aline.Vidotto@tcd.ie}, V.~Bourrier$^2$
\\ 
$^{1}$School of Physics, Trinity College Dublin, the University of Dublin, Dublin-2, Ireland\\
$^{2}$Observatoire de l'Universit\'e de Gen\`eve, Chemin des Maillettes 51, Versoix, CH-1290, Switzerland
}

\date{Accepted XXX. Received YYY; in original form ZZZ}

\pubyear{2017}

\begin{document}
\label{firstpage}
\pagerange{\pageref{firstpage}--\pageref{lastpage}}
\maketitle

\begin{abstract}
Winds of cool dwarfs are difficult to observe, with only a few M dwarfs presenting observationally-derived mass-loss rates ($\mdot$), which span several orders of magnitude. Close-in exoplanets are conveniently positioned in the inner regions of stellar winds and can, thus, be used to probe the otherwise-unobservable local properties of their host-stars' winds. Here, we use local stellar wind characteristics observationally-derived in the studies of atmospheric evaporation of the warm-neptune \gj b to derive the global characteristics of the wind of its M-dwarf host. Using an isothermal wind model, we constrain the stellar wind temperature to be in the range [0.36,0.43] MK, with $\mdot=[0.5,2.5] \times 10^{-15} ~\msano$. By computing the pressure balance between the stellar wind and the interstellar medium, we derive the size of the astrophere of \gj\ to be around 25~au, significantly more compact than the heliosphere. We demonstrate in this paper that transmission spectroscopy, coupled to planetary atmospheric evaporation and stellar wind models, can be a useful tool for constraining the  large-scale wind structure of planet-hosting stars. Extending our approach to future planetary systems discoveries will open new perspectives for the combined characterisation of planetary exospheres and winds of cool dwarf stars. 
\end{abstract}

\section{Introduction}\label{sec.intro}
Stellar winds play important roles in stellar rotational evolution \citep{1962AnAp...25...18S, 1958ApJ...128..664P,1967ApJ...148..217W} and exoplanetary systems \citep{2004ApJ...602L..53I,2009ApJ...703.1734V,2013MNRAS.436.2179L,2005A&A...437..717G,2010ApJ...720.1262V,2007AsBio...7..167K,2007AsBio...7..185L,2010Icar..210..539Z}. Unfortunately, the tenuous winds of cool dwarf stars (F to M) are very challenging to be measured \citep[e.g.,][]{2004LRSP....1....2W}. So far, only a few techniques have been used to probe these rarefied winds, which implies that properties such as terminal (asymptotic) wind velocity, mass-loss rates and temperatures, are still restricted to about a dozen or so low-mass stars \citep{2011ApJ...741...54C}.

Some attempts to detect winds from dwarf stars have been inspired on the theory of radio free-free thermal emission from the winds of evolved cool stars \citep{1975A&A....39....1P}. Through the lack of detection of radio emission, several works {have} placed important upper limits on mass-loss rates of cool dwarf stars \citep[e.g.,][]{1996ApJ...462L..91L,1997AandA...319..578V,2000GeoRL..27..501G,2014ApJ...788..112V, 2017A&A...599A.127F,2017A&A...602A..39V}. Another suggested method proposed to probe stellar winds is through the detection of X-ray emission generated when ionised wind particles {charge exchange} with neutral atoms of the interstellar medium \citep{2002ApJ...578..503W}. Similarly to the radio technique, this technique has so far only provided upper limits on mass-loss rates of dwarf stars. A third method, which has been used more often, lies on modelling the observed \ly\ line profile of the target star, taking into consideration the absorption of the line profile along the line-of-sight \citep{2001ApJ...547L..49W}. Absorption occurs as the \ly\ photons travel through the stellar astrosphere\footnote{The astrosphere is the stellar equivalent of the heliosphere, i.e., it is the region surrounding the star that is permeated by the stellar wind. The outer boundary of the astrosphere is the interstellar medium.}, the interstellar medium and the heliosphere. The reconstruction of the \ly\ line, through a combination of hydrodynamical modelling and high-spectral resolution observations, has enabled estimates of mass-loss rates for about a dozen dwarf stars \citep{2004LRSP....1....2W}.

These three aforementioned techniques were used in a few attempts to measure the mass-loss rates of the closest star to our Sun, Proxima Centauri. Only  upper  limits were derived  for Prox Cen's mass-loss rate (Table \ref{table0}). This lack of detection of the wind of the closest star to our Sun stresses the challenge of measuring wind properties of cool dwarf stars and, in particular, of M dwarf (dM) stars, such as Prox Cen. In addition to Prox Cen, estimates of mass-loss rates also exist for other dM stars, in particular those members of eclipsing binary systems, in which the primary is a white dwarf. In these systems, the white dwarf accretes, presumably through a Bondi-Hoyle type accretion, the stellar wind material from the dM companion \citep{2006ApJ...652..636D,2012MNRAS.420.3281P}. By modelling the accretion rates, the stellar wind outflow rates can be inferred. {At the moment, it is unclear if the coronae and winds of dM stars in these binary systems are similar to those of isolated/non-interacting dM stars.}
 Table \ref{table0} summarises the mass-loss rates inferred for dM stars, {where the dM stars in close binary systems are marked with an asterisk. For the interacting stars where only mass-accretion rates onto the white dwarf companions have been reported \citep{2012MNRAS.420.3281P}, these values are set as lower limits for the mass-loss rates of the dM stars. For example, \citet{2006ApJ...652..636D} finds that dM mass-loss rates are about 15 to 100 times larger than white dwarf mass-accretion rates.}

\begin{table*}
\caption{Mass-loss rates inferred for M dwarf stars.} \label{table0}
\begin{center}
\begin{tabular}{llrllllcccccccccc}
\hline
Star & sp. & $\mdot$ & $\mdot$ & reference & $\log(L_X)$ & reference\\
 &type& $(\msano)$ & $(\mdot_\odot)$ & ($\mdot$) &(erg~s$^{-1}$)& ($L_X$)\\ \hline
\gj & M2.5 & $(0.45 - 2.5)\e{-15}$ & $0.02-0.13$& This work & $26.76$ & \citet{2015Natur.522..459E}\\ \hline
EV Lac & M3.5 & $2\e{-14}$& $1$ & \citet{2005ApJ...628L.143W} &$28.99$& \citet{2005ApJ...628L.143W}\\ 
SDSS J1212-0123 &M4$^*$ & {$> 6.4\e{-17}$} & $> 0.003$ & \citet{2012MNRAS.420.3281P}& - & - \\
WD 1026$+$002 & M4$^*$ & $1\e{-16}$ & 0.005 & \citet{2006ApJ...652..636D} & - & -\\
EG UMa & M4$^*$ &  $1\e{-16}$ & 0.005 & \citet{2006ApJ...652..636D} & - & -\\ 
GK Vir &M4.5$^*$& {$>2.2\e{-17}$} & $>0.001$ & \citet{2012MNRAS.420.3281P} & - & -\\
YZ CMi & M4.5& $<1\e{-12}$ & $<50$ & \citet{1996ApJ...462L..91L}&$28.55$&\citet{2008MNRAS.390..567M}\\
 \ldots &\ldots &$<2\e{-12}$ & $<100$ & \citet{1997AandA...319..578V}&\ldots&\ldots \\
Prox Cen & M5.5 & $<7\e{-12}$ & $<350$ & \citet{1996ApJ...460..976L}&$27.22$& \citet{2005ApJ...628L.143W} \\ 
\ldots &\ldots & $<4\e{-15}$ & $<0.2$ & \citet{2001ApJ...547L..49W} &\ldots&\ldots \\
\ldots &\ldots & $< 2.8\e{-13}$ & $<14$ & \citet{2002ApJ...578..503W} &\ldots& \ldots \\
RR Cae & $\gtrsim$ M6.5$^*$ &  $6\e{-15}$&0.3 & \citet{2006ApJ...652..636D} & - & -\\ 
\hline
\end{tabular}
\end{center}
         $^*$Secondary star of eclipsing white dwarf $+$ M dwarf binary.
\end{table*}

Another indirect way to probe stellar winds, and that will be the focus of this paper, is by a relatively recent, and promising, technique used in studies of exoplanetary evaporation \citep[][B16, from now on]{2008Natur.451..970H,2010ApJ...709..670E,2010ApJ...709.1284B, 2013A&A...557A.124B,2014Sci...346..981K, 2016A&A...591A.121B}. Through modelling of the interaction between the upper planetary atmosphere and the stellar radiation pressure and winds, the reconstruction of the transmitted spectrum of the star  through the atmosphere of the transiting planet is capable of placing constraints on the evaporation (mass-loss rates and velocities) of exoplanetary atmospheres. As a by-product of this technique, one can then place constraints on the local conditions of the wind of planet-hosting stars. In this technique, the planet is used as a probe for the stellar wind, i.e., a sort of {\it wind-ometer}. 

The characterisation of winds of dM stars is  also of strong relevance for studies of planetary habitability. dM stars have been central in the search for potentially habitable terrestrial planets -- their large numbers in the Galaxy, small radii, and closer-in habitable zones makes them the prime targets to detect exoplanets  orbiting within this region \citep[e.g.][]{
2017Natur.542..456G,2017arXiv170405556D}. A planet orbiting in the habitable zone could retain liquid water on its surface \citep[e.g.,][]{1993Icar..101..108K}. However,  the presence of a planet in a habitable zone is not an indicator of habitability, as many other factors should be taken into consideration \citep[e.g., ][]{2007AsBio...7...85S,2007AsBio...7...30T,2009A&ARv..17..181L}, including the presence of stellar winds, which can erode an unprotected planetary atmosphere,  being harmful for the creation and development of life \citep{2010Icar..210..539Z,2011MNRAS.412..351V,2013A&A...557A..67V,2014A&A...570A..99S,2016ApJ...820L..15D}. 

In this paper, we focus on the wind of the planet-hosting star \gj , an M2.5 dwarf star. The goal of this paper is to use results provided by studies of exoplanetary evaporation to obtain additional constraints on the wind of \gj . These constraints include: mass-loss rates,  wind temperature, wind characteristics as a function of distance to the star, terminal velocities and astrospheric sizes. This paper is organised as follows. Section \ref{sec.observations} summarises previous local constraints derived for the wind of \gj , which are then used to feed our global stellar wind models, presented in Section \ref{sec.model}. Our results are shown in Section \ref{sec.results}, followed by our discussion and conclusions in Sections \ref{sec.discussion} and \ref{sec.conclusions}.

\section{Previous constraints derived for  the wind of \gj}\label{sec.observations}
\begin{figure}  
\includegraphics[width=\columnwidth]{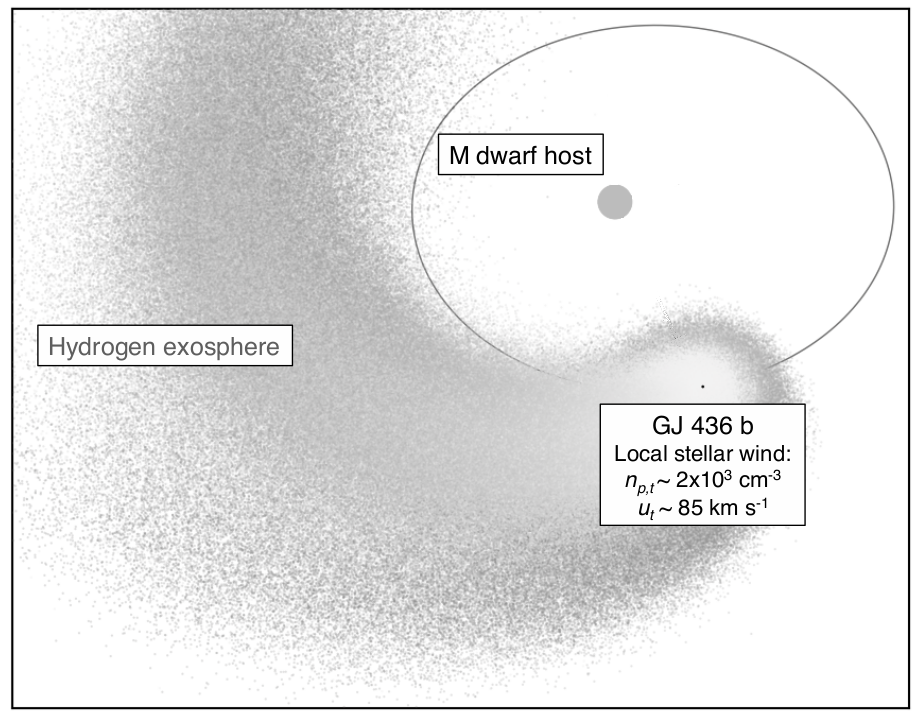}
\caption[]{Schematic representation of the neutral hydrogen cloud surrounding \gj b. Planetary orbit is shown by the ellipse. Figure adapted from B16.}
\label{fig.cloud_schematics}
\end{figure}

Observations of the warm-neptune \gj b performed with HST/STIS at three different epochs in the stellar \ly\ line revealed deep, repeatable transits attributed to a giant exosphere of neutral hydrogen \citep{2014ApJ...786..132K,2015Natur.522..459E,2015A&A...582A..65B,2016A&A...591A.121B}. \citet{2015A&A...582A..65B} and \citetalias{2016A&A...591A.121B} explained the structure of this exosphere by the combined effects of radiative braking and stellar wind interactions on the neutral hydrogen atoms escaping the planet upper atmosphere (see Figure \ref{fig.cloud_schematics}). While the low radiation pressure from the dM host star brakes the gravitational fall of the escaping gas and allows its dispersion within a large volume around the planet, charge-exchange between the escaping gas and stellar wind protons leads to the abrasion of the outer regions of the exosphere, and the partial neutralisation of the wind. This secondary population of neutralised protons has the velocity distribution of the stellar wind, which differs from the dynamics of the planetary neutrals that originally escaped from the planet. 
{This interpretation has been strengthened by new observations of \gj\ in the \ly\ line (Lavie et al.~2017, in prep.), which completed the coverage of the exospheric transit up to about 20h after the transit of the planet, and validated the structure predicted by \citetalias{2016A&A...591A.121B} for the exospheric tail.} Absorption of the stellar \ly\ line by the exosphere of \gj b thus traces both the properties of the planetary upper atmosphere and that of the stellar wind.\footnote{Although the astrosphere, the interstellar medium, the heliosphere and the exoplanetary atmosphere can all absorb the \ly\ line, only the latter causes variations on short timescales (a few hours during transits), while variations of the stellar \ly\ line occur on longer timescales (several years).} 

 Given the complexity of extended exospheres, such properties are best retrieved by the use of three-dimensional (3D) numerical models, such as with the EVaporating Exoplanets (EVE) code. The code EVE was developed to calculate the structure of an exoplanet upper atmosphere and the transmission spectrum of the species it contains, like neutral hydrogen \citep[e.g.,][]{2013A&A...557A.124B,2015A&A...582A..65B}.  EVE performs Monte-Carlo particle simulations to compute the dynamics of hydrogen meta-particles in the exosphere. Particles are subjected to a number of physical mechanisms, including radiation pressure and interactions with the stellar wind. The code calculates theoretical \ly\ spectra as seen with the HST after absorption by the planet, its atmosphere, and the interstellar medium. The direct comparison of these spectra with observational data allows for the inference of several physical properties of both the planetary atmosphere and the star. EVE has been used in this way to study the exospheres of most known evaporating exoplanets \citep[e.g.,][]{2013A&A...557A.124B,2012A&A...547A..18E,2014A&A...565A.105B,2015A&A...582A..65B}. 
 
\citetalias{2016A&A...591A.121B} used the code EVE  to simulate the exosphere of \gj b. These authors were able to place strong constraints on the bulk velocity, kinetic dispersion and density of the stellar wind protons, described with a Maxwellian distribution, at the location of  \gj b. By independently fitting the HST observations secured at different epochs, \citetalias{2016A&A...591A.121B} found that the bulk velocity of the stellar wind is stable over time with a value $85^{+6}_{-16}$ km s$^{-1}$. Small deviations to the stable structure of the exosphere detected between the different epochs were attributed to variations in the stellar wind density, which was found to be in the range [0.9, 4.8]$\e3$ cm$^{-3}$, computed at the semi-major axis distance $a$. Table \ref{table1} shows the physical properties of the system and Table \ref{table2} compiles the relevant properties of the wind of \gj . The proton density presented in Table \ref{table2} is the one computed at at mid-transit (\citetalias{2016A&A...591A.121B} present values computed at $a$ instead)  and is in the range [0.8,4.2]$\e3$ cm$^{-3}$.  Finally, we note that \citetalias{2016A&A...591A.121B} derived an upper limit of about $20$ km s$^{-1}$ on the {3D} kinetic dispersion of the wind protons around the bulk motion of the wind. As in the present paper we use a (1D) hydrodynamical approach to describe the stellar wind, our model is not sensitive to these random (thermal) velocities of the particles. 

\begin{table}
\caption{Properties of the \gj\ planetary system.} \label{table1}
\begin{center}
\begin{tabular}{llllllccccccccccc}
\hline
stellar mass$^a$ & $M_\star$ & 0.452& $M_\odot$  \\
stellar radius$^b$ & $R_\star$ & 0.437& $R_\odot$   \\
age$^a$ & $\tau$ & $6^{+4}_{-5}$& Gyr    \\
X-ray luminosity$^c$ & $L_X$ & $5.7\e{26}$& erg s$^{-1}$ \\
semi-major axis$^b$ & $a$ & 0.0287& au \\
 \ldots &  $a$ & $14.1$ & $R_\star$    \\
Eccentricity$^d$ & e & 0.16 & &  \\  
mid-transit orbit & $a_t$ & $0.0306$ & au \\
\ldots & \ldots & $15.1$ & $R_\star$ \\
distance & $d$ & $10.14$ & pc \\
         \hline
\end{tabular}
\end{center}
$^a$\citet{2007ApJ...671L..65T};
$^b$\citet{2011ApJ...735...27K};
$^c$\citet{2015Natur.522..459E};
$^d${\citet{2014AA...572A..73L}}
\end{table}

\begin{table}
\caption{Local and global stellar wind properties of \gj . The local  velocity and proton density  represent the local stellar wind conditions at the position of \gj b during mid-transit (represented by the subscript ``t'').} \label{table2}
\begin{center}
\begin{tabular}{llllllcccccccc}
\hline
local velocity$^a$ & $u_t$& $85^{+6}_{-16}$ & km s$^{-1}$  \\
local proton density$^a$ & $n_{p,t}$ & $2.0^{+2.2}_{-1.2}\e3$ & cm$^{-3}$ \\
mass-loss rate$^b$ & $\mdot$ & $1.2^{+1.3}_{-0.75} \times 10^{-15}$&$\msano$ & \\
\ldots & \ldots & $0.059^{+0.074}_{-0.040}$ & $\mdot_\odot$  \\    
wind temperature$^b$& $T$ & $0.41^{+0.02}_{-0.05}$& MK \\
terminal velocity$^b$ & $u_\infty$ & $370^{+10}_{-30}$ & km s$^{-1}$\\
ISM column density$^c$ & $N_{\rm ISM}$ & $1\e{18}$ & cm$^{-2}$  \\
ISM average H density$^b$ & $n_{\rm ISM}$ & $0.03$ & cm$^{-3}$ \\
ISM/star rel.~velocity$^b$ & $v_{\rm ISM}$ & $81$ & km s$^{-1}$ \\
astrospheric size$^b$ & $R_{\rm ast}$ & 25 & au  \\
 \hline
\end{tabular}
\end{center}
$^a$\citetalias{2016A&A...591A.121B}; $^b$This work; $^c$\citet{2015A&A...582A..65B}
\end{table}


\section{The stellar wind model }\label{sec.model}
Through the modelling of the exospheric escape of \gj b with the code EVE, \citetalias{2016A&A...591A.121B} determined the local stellar wind velocity and density. This, as a consequence, advances our understanding of the {\it local} interplanetary medium. However, to be able to characterise the {\it global} characteristics of the wind of the host star, we need to couple these results to a wind model. In the present work, we assume use a spherically symmetric, steady-state isothermal wind model \citep{1958ApJ...128..664P} to derive the global characteristics of the wind of \gj . For a hot coronal wind (i.e., with temperatures $T$ of around a million K), thermal pressure gradient is able to drive the wind. The wind radial velocity $u$ is thus derived   from the integration of the momentum equation
\begin{equation}  \label{eq.parker}
\rho u \frac{\partial u}{\partial r} = -\frac{\partial }{\partial r} \left(\frac{\rho}{\mu m_p} k_B T\right) - \rho \frac{G M_\star}{r^2} , 
\end{equation} 
where $\rho = n \mu m_p $ is the wind mass density, $\mu$  the mean-particle weight, $m_p$ the proton mass, $r$  the radial coordinate,   and $k_B$ and $G$ are  the Boltzmann and gravitational constants. Here, we assume that the stellar wind is a fully ionised hydrogen plasma ($\mu=0.5$), so that its total particle density is $n=n_p/\mu=2n_p$. The equation of motion (\ref{eq.parker}) can be integrated to yield \citep{1999isw..book.....L}
\begin{equation}  \label{eq.parker_sol}
u \exp \left(- \frac{u^2}{2 c_s^2} \right) = c_s \frac{r_s^2}{r^2} \exp \left( \frac32 -\frac {2r_s}{r} \right)
\end{equation} 
where $c_s=(k_B T/\mu m_p)^{1/2}$ is the isothermal sound speed and $r_s = {GM_\star (2c_s^2)^{-1}}$ is the point beyond which the solution becomes supersonic. Equation (\ref{eq.parker_sol}) is a transcendental equation that can be easily solved numerically for $u$ provided the wind temperature $T$ and stellar mass $M_\star$ and radius $R_\star$ are given (see also \citealt{2004AmJPh..72.1397C}). 

Under the assumption of a spherically symmetric stellar wind, mass conservation implies that the wind mass loss rate is given by $\mdot = 4 \pi r^2 \rho u  = \rm{constant} $. Therefore, the wind density profile $\rho(r)$ is
\begin{equation} \label{eq.rho}
\rho = \frac{\mdot}{ 4 \pi r^2 u } \, ,
\end{equation} 
with $u$ given by Equation (\ref{eq.parker_sol}).

\section{Global characteristics of the wind of \gj }\label{sec.results}
Using the model presented in Section \ref{sec.model} and the observational constraints derived for the local conditions of the wind of \gj\ (\citetalias{2016A&A...591A.121B}, i.e., at the mid-transit orbital distance), we derive next the global characteristics of the wind of \gj , namely the wind mass-loss rate, temperature and the velocity and density profiles.

Mass conservation implies that 
\begin{equation}  \label{eq.mdot}
\mdot = 4 \pi r^2 \rho u = 4 \pi a_t^2 (n_{p,t} m_p) u_t \, ,
\end{equation} 
where the subscript ``t'' indicates that the quantities are calculated at the position of the planet at mid-transit. From the values derived by \citetalias{2016A&A...591A.121B} (cf. Table \ref{table2}), we find a mass-loss rate for \gj\ of $1.2^{+1.3}_{-0.8} \times 10^{-15} ~\msano$. 

\begin{figure}
	\includegraphics[width=0.47\textwidth]{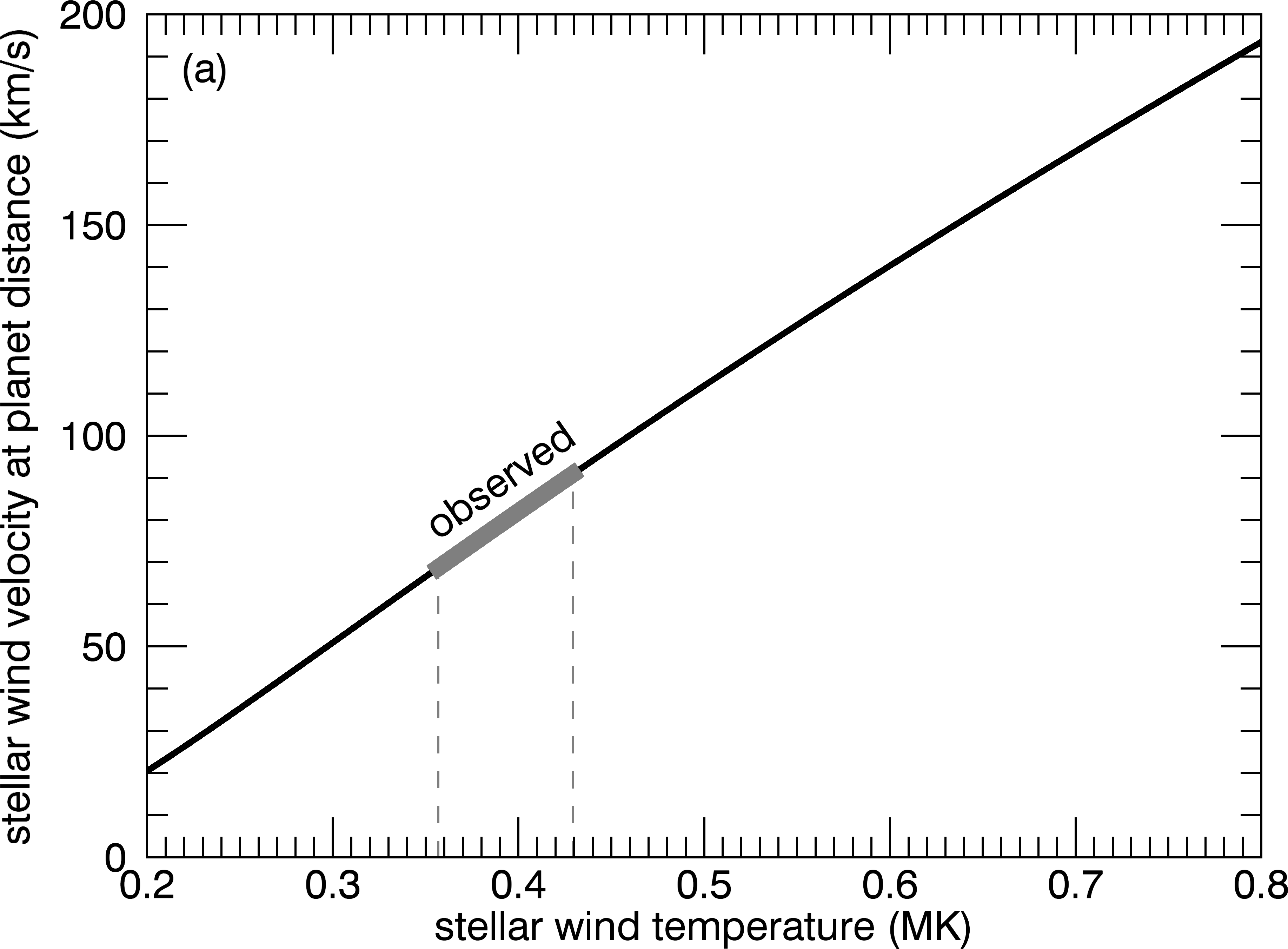}\\
	\includegraphics[width=0.47\textwidth]{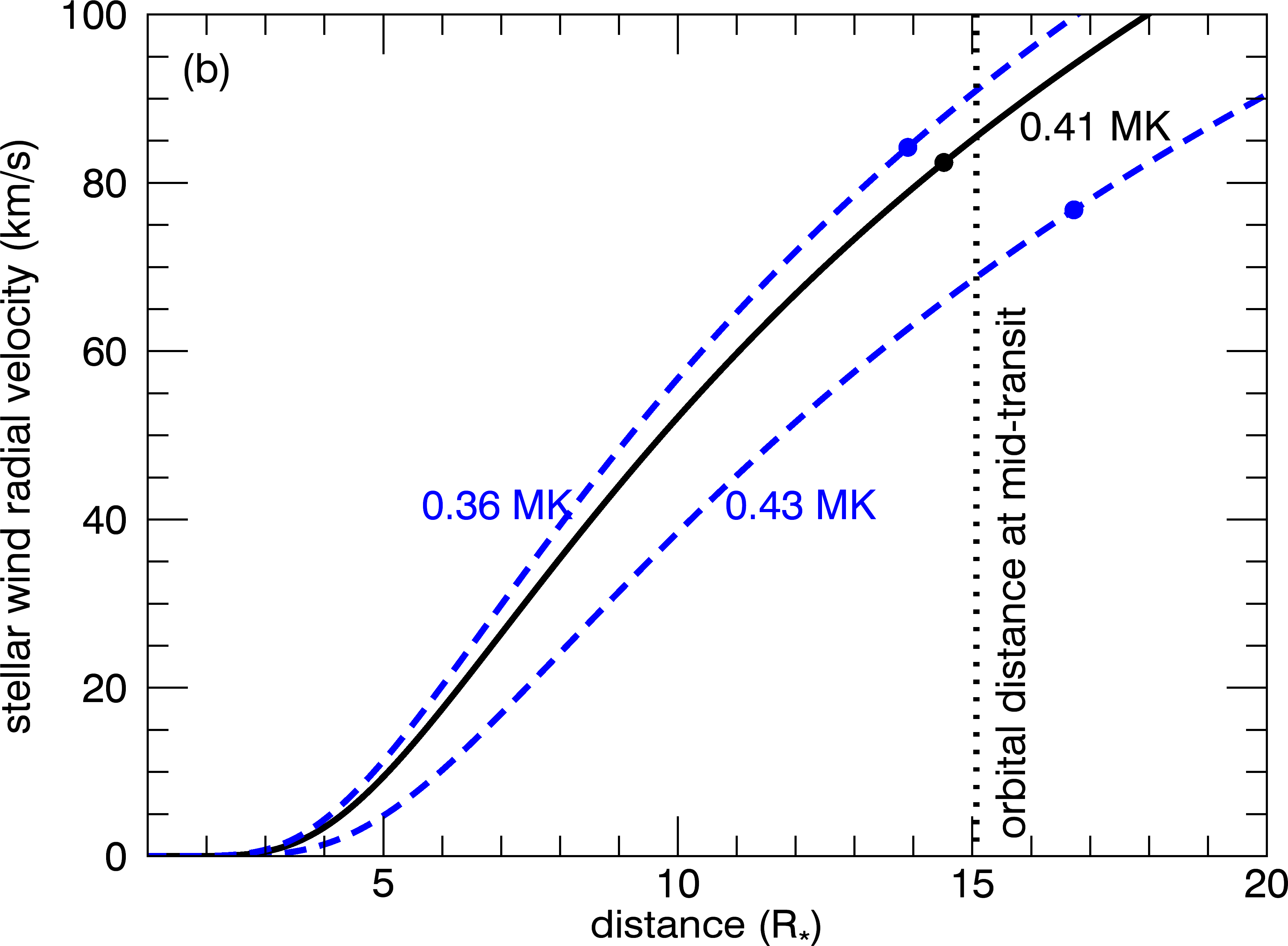}\\
\caption{(a) Model-predicted local wind velocity (i.e., at the position of the planet at mid-transit) versus the wind temperature. The observationally-derived velocity of $85^{+6}_{-16}$~km~s$^{-1}$ at the position of the planet (cf. Table \ref{table1}, \citetalias{2016A&A...591A.121B}) constrains the stellar wind temperature of \gj\ to be in  the range between $0.36$ and $0.43$~MK. (b) The inner velocity profile of the stellar wind. Black solid line assumes a wind temperature that best matches the observationally-derived velocity at the planet orbit, while the blue dashed lines are associated to the 1-$\sigma$ uncertainty in the observed velocity. Circles indicate the distance at which the wind becomes supersonic.}\label{fig.wind_temperature}
\end{figure}

Since $T$ is an unconstrained quantity of the stellar wind, we use the observationally-derived {bulk} velocity at the position of the planet (cf. Table \ref{table2} and \citetalias{2016A&A...591A.121B}) to derive the wind temperature. We start by solving Equation (\ref{eq.parker_sol}) for a range of wind temperatures from $0.2$ to $0.8$~MK. The top panel of Figure \ref{fig.wind_temperature} shows the wind velocity at the position of the planet  as a function of the adopted wind temperature. The wind model that best matches the observationally-derived velocity of $85^{+6}_{-16}$~km s$^{-1}$ has a temperature of $0.41^{+0.02}_{-0.05}$~MK (i.e., temperatures in the range of $0.36$ to $0.43$~MK).\footnote{{It is worth comparing the velocity of the wind of \gj\ with those of the solar wind. At a distance of $15~R_\odot$, which is the equivalent normalised distance to \gj b's orbit, the observed velocities of the solar wind range from  175 to 350 km~s$^{-1}$, which is consistent to a 1-MK isothermal  solution for the solar wind \citep{2000JGR...10525133W}. Compared to the wind of \gj , the solar wind has both higher local velocities and modelled temperature at the equivalent distance to \gj b. }} {Using the empirical relation between the average coronal temperature and X-ray fluxes from \citet{2015A&A...578A.129J}, the predicted average coronal temperature of \gj\ is about 1.8~MK. This is larger than the temperature predicted by our wind model.  This difference can be explained by the different origins of X-ray emission and stellar wind. As it happens in the Sun, it is likely that in \gj , the X-ray emission arises from the closed-magnetic field line region, while the wind originates in open field line regions (coronal holes), which are X-ray dark. Our low temperature for \gj\ might thus indicate that the closed coronal region is hotter than the open-magnetic field line region.}


The bottom panel of Figure \ref{fig.wind_temperature} shows the velocity profiles of the stellar wind for three different wind temperatures. The  black solid line is for a wind temperature of $0.41$~MK (best match of the observationally-derived velocity at the planet orbit) and the blue dashed lines for temperatures of $0.36$ and $0.43$~MK (i.e., associated to the 1-$\sigma$ uncertainty in the observed velocity). The sonic point (circles in Figure \ref{fig.wind_temperature}b),  i.e., the point beyond which the wind velocity becomes supersonic, is, by coincidence, close to the orbital distance of \gj b. Figure \ref{fig.wind_temperature}b  shows that, depending on the temperature of the wind, the planet might orbit in the subsonic wind. In spite of this, the motion of \gj b through the stellar wind is likely supersonic, as the Keplerian velocity of the planet is $\sim 120$~km~s$^{-1}$. 

Although the transit observations allow us to infer the density $n_{p,t}$ at the position of the planet, for us to obtain the density at wind base $ n_0$, we need to solve the wind equations, as a simple $r^{-2}$ law only works in the asymptotic wind regime (i.e., far from the star, where the velocity approaches $u_\infty$). From Equation (\ref{eq.mdot}), we have that
\begin{equation}  \label{eq.mdot2}
  n_{p,0}   = \left( \frac{a_t}{R_\star} \right)^2 \frac{u_t}{u_{0}} n_{p,t}  \, ,
\end{equation} 
where the subscript ``0'' indicates quantities evaluated at the wind base, which we take to be at $r=R_\star$. The wind velocity at the wind base $u_{0}$ is obtained from the solution of the momentum equation (\ref{eq.parker}), while $u_t$, $n_{p,t}$, $a_t$ and $R_\star$ are given in Tables \ref{table1} and \ref{table2}. 

\begin{figure}
	\includegraphics[width=0.47\textwidth]{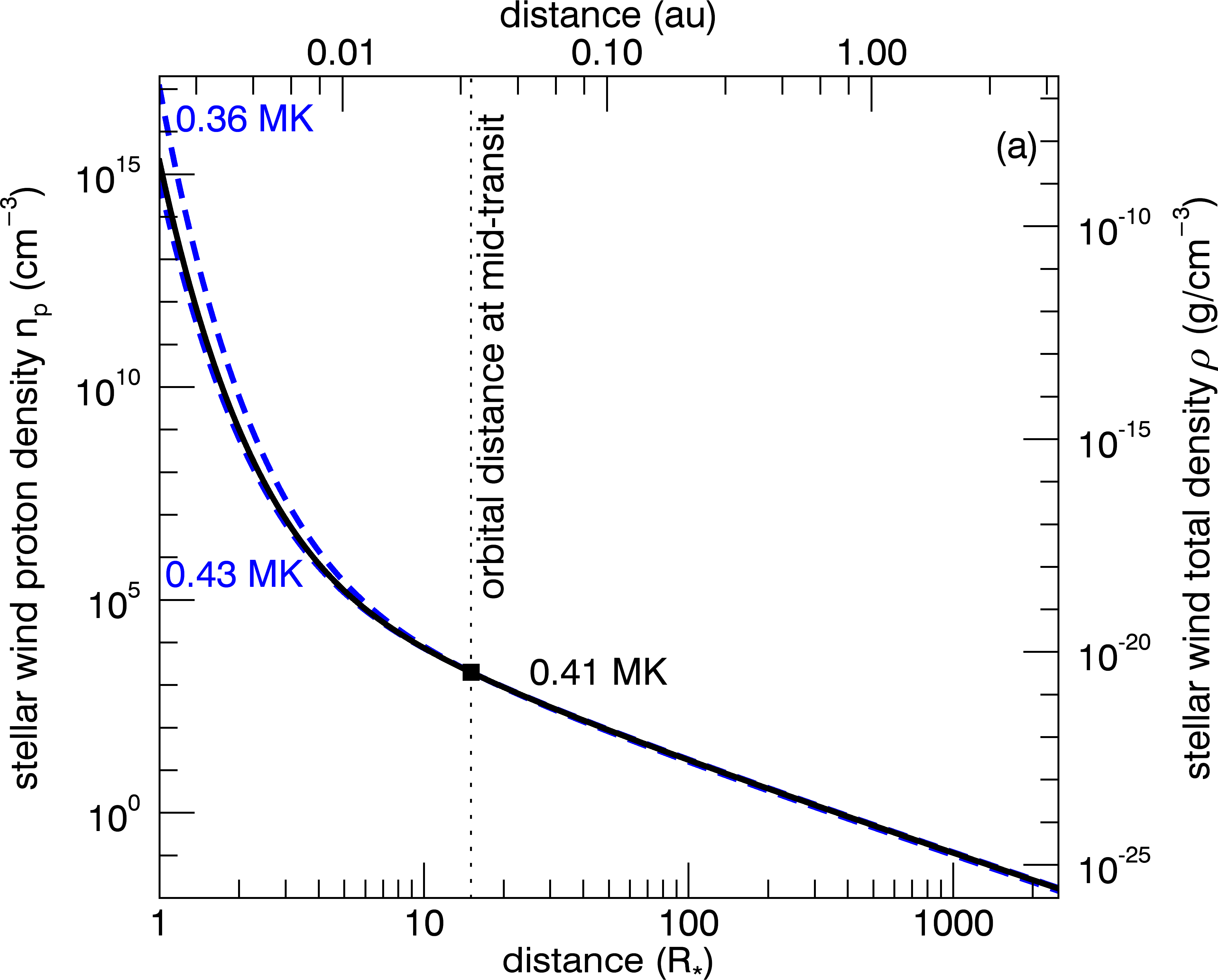}
	\includegraphics[width=0.47\textwidth]{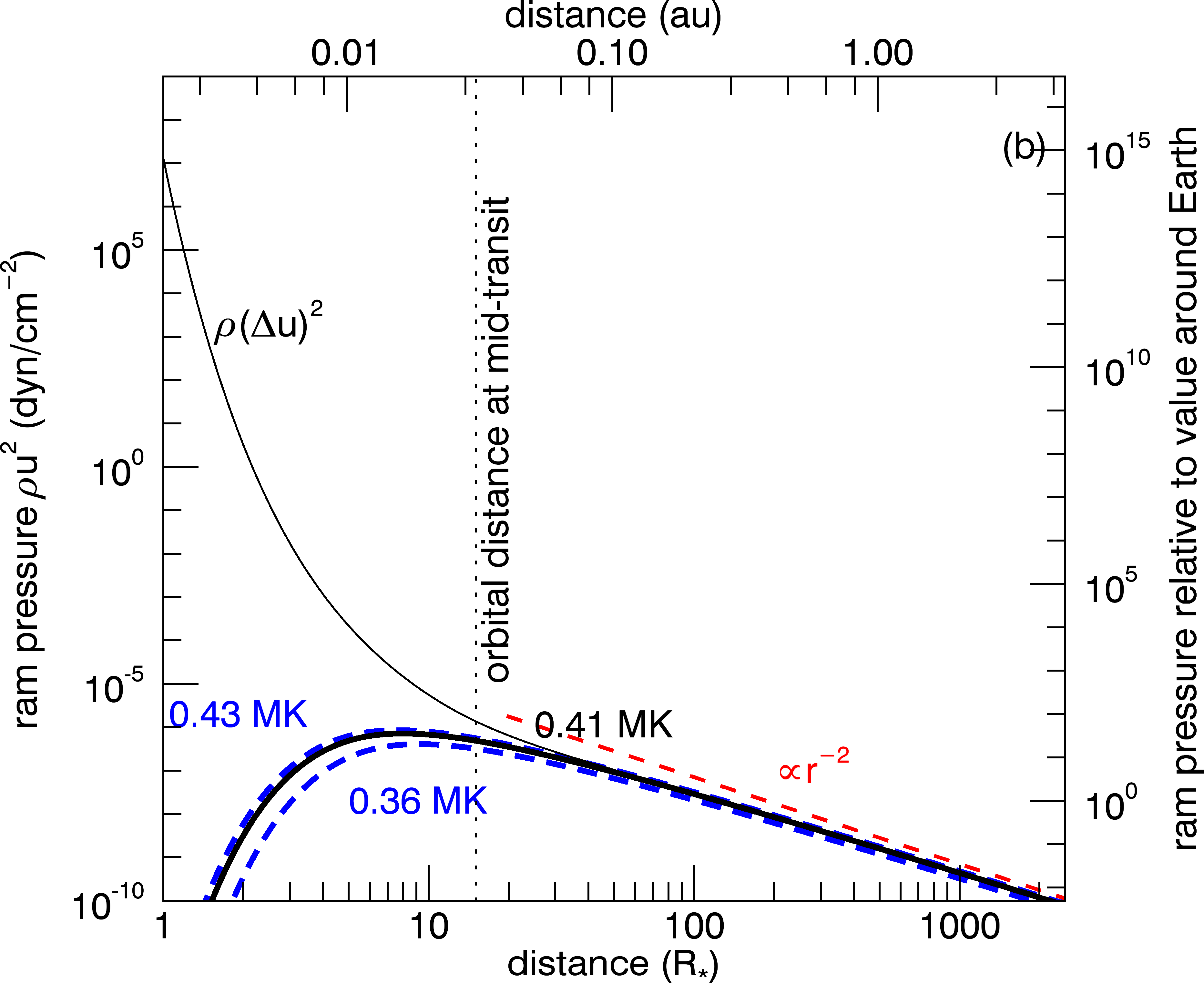}
\caption{(a) Stellar wind density profile of \gj\  as a function of distance for different wind temperatures. The vertical left (right) axis shows the proton (total mass) density. The solid lines are as in Figure \ref{fig.wind_temperature}b and were scaled to match the observationally-derived local wind density  at the position of the planet (square). (b) The same, but for the stellar wind ram pressure {calculated according to Equation (\ref{eq.pram}). The thin black line represents the ram pressure as seen by an orbiting planet in circular Keplerian motion. See text for more details.} }\label{fig.wind_dens}
\end{figure}

Figure \ref{fig.wind_dens}a shows the density profile of the stellar wind, where the left (right) axis shows the proton (total mass) density. The solid lines are as in Figure \ref{fig.wind_temperature}b, where the black solid line is the solution for the case with a wind temperature of  $0.41$~MK (best match of the observationally-derived velocity at the planet orbit), and the blue dashed lines represent wind temperatures associated to the 1-$\sigma$ uncertainty in the observationally-derived velocity. The density profiles shown in Figure \ref{fig.wind_dens} were scaled to match the local wind density at the position of the planet (square). The model with a wind temperature of  $0.41$~MK predicts a stellar wind base densities of about $2 \times 10^{15}$~cm$^{-3}$. The base density decreases by a factor of 3 for a wind temperature of $0.43$~MK and increases by a factor of $\sim$ 50 for $0.36$~MK, showing that the density profile is very sensitive to wind temperature. The  base density is likely overestimated in our models, but the density profile is less affected in the super-sonic region of the wind, as we discuss in Section \ref{sec.discussion}. 


Our modelling also allows us to compute the ram pressure of the wind, which is required to calculate the extension of planetary magnetopauses, in the case of a planets with intrinsic magnetospheres  \citep[e.g.,][]{1930Natur.126..129C}, or the extension of ionopauses, in the case of direct interaction of the stellar wind with ionospheres \citep[e.g.][]{1986SSRv...44..241L}. {It is also used to calculate the size of astrospheres, as we present further below. The stellar wind ram pressure is given by
\begin{equation}\label{eq.pram}
P_{\rm ram} = \rho {\bf u}^2
\end{equation}
where $\rho$ is the mass density and ${\bf u}$ is the wind velocity. Figure \ref{fig.wind_dens}b shows the ram pressure profile as a function of distance for the wind of \gj . For planets orbiting at close distances to their host stars, the planet's orbital Keplerian velocity  $u_K$ might be comparable to or even larger than the local stellar wind velocity \citep[e.g.][]{2010ApJ...722L.168V}. In these cases, the ram pressure exerted on the planet is given by $P_{\rm ram} = \rho (\Delta {\bf u} )^2$, where $\Delta {\bf u} = {\bf u} - {\bf u_K}$ is the relative velocity of the planet through the stellar wind. This is represented in Figure \ref{fig.wind_dens}b by the thin black line. As we notice, for planets orbiting at distances beyond $20~R_\star$, the Keplerian velocity becomes negligible compared to the wind velocity and $P_{\rm ram}$ approaches the solution from Equation (\ref{eq.pram}). The ram pressure exerted on \gj b, which includes the Keplerian velocity of $120$~km~s$^{-1}$, is $1.4\e{-6}$~dyn~cm$^{-2}$. Assuming \gj b has a dipolar magnetic field, an upper limit for \gj b's magnetic field of 60\% Jupiter's value is inferred \citepalias{2016A&A...591A.121B}. In Figure \ref {fig.wind_dens}b,} the ram pressure we derive is also shown relatively to the value in the solar wind ram pressure in the vicinity of the Earth ($P_{{\rm ram,}\oplus}\simeq 2\e{-8}~$dyn~cm$^{-2}$, cf.~right axis in Figure \ref {fig.wind_dens}b). The stellar wind of \gj\ reaches the value of $P_{{\rm ram,}\oplus}$ at a distance of $\sim 0.25$~au, or about $123~R_\star$, which is much closer in than the equivalent distance at the solar system. 

{At large distances, when the wind velocity has reached terminal speed, the ram pressure falls off with $r^{-2}$ (cf.~red dashed line in Figure \ref{fig.wind_dens}b).} As the star travels through the interstellar medium (ISM), the ram pressure of the stellar wind is balanced by the ram pressure of the interstellar ``wind'' ($P_{\rm ISM}$, e.g.,  \citealt{1998ApJ...492..788W}). Shock formation occurs, thus, in the upwind direction, delimiting the size of the astrophere. As $P_{\rm ram} \propto r^{-2}$, the stand-off distance to the termination shock can be estimated as
\begin{equation}\label{eq.pism}
R_{\rm ast} = \left( \frac{P_{\rm ram} (r_1)}{P_{\rm ISM}} \right)^{1/2} r_1 = \left( \frac{P_{\rm ram} (r_1)}{m_p n_{\rm ISM}  v_{\rm ISM}^2} \right)^{1/2} r_1 ,
\end{equation}
where $r_1$ is a reference radius within the regime obeying the $r^{-2}$ power-law decay, $n_{\rm ISM}$ is the  ISM density of neutral hydrogen and $v_{\rm ISM}$ is the ISM velocity as seen by \gj . 
Assuming the ISM is homogeneous along the line-of-sight toward \gj\  {and purely neutral}\footnote{{The local ISM is partly ionised, so the assumption that the ISM is purely neutral is a crude one and gives a lower limit for the ISM density. Consequently, the astrospheric size we derive is to be considered an upper limit.}}, we estimate  $n_{\rm ISM} = N_{\rm ISM} /d =  0.03 {\rm ~cm}^{-3}$, where the distance is $d = 10.14$~pc and $N_{\rm ISM} =10^{18}$~cm$^{-2}$ is the  ISM column density derived in \citet{2015A&A...582A..65B}. These authors  identified the ISM cloud along the line-of-sight toward \gj\ as being most consistent with the Local Interstellar Cloud (LIC). The heliocentric LIC flow vector in the direction of \gj\ is derived using the Local ISM Kinematic Calculator \citep{2008ApJ...673..283R} and the radial velocity of LIC derived in \citet{2015A&A...582A..65B}. The heliocentric velocity of \gj\ is calculated using its Hipparcos proper motion \citep{2007A&A...474..653V} and its radial velocity \citep{2015A&A...582A..65B}. With these, we obtain a relative velocity between the ISM and \gj\ of $v_{\rm ISM} = 81$~km~s$^{-1}$ and a ram pressure of $P_{\rm ISM}\sim 3.3 \e{-12}~$dyn~cm$^{-2}$. From Equation (\ref{eq.pism}) and Figure \ref{fig.wind_dens}b, we find that the distance to the termination shock in the astrosphere {of \gj\ is $\sim 1.2 \times 10^4~R_\star$, which is about $25$~au.}


\section{Discussion}\label{sec.discussion}

Close-in planets are conveniently located in the stellar wind acceleration zone. By studying the interactions between planets and the winds of their host stars, these planets can be used as a much-needed additional tool for probing winds of dwarf stars. Transit spectroscopy together with atmospheric escape modelling can provide {\it local} velocities, densities and temperatures, which can then inform wind models, allowing us to more accurately derive the global characteristics of the stellar wind. 

We used the  {\it local} velocity and density determined in a previous study of planetary evaporation by \citetalias{2016A&A...591A.121B} to calculate the mass-loss rate of \gj . The assumption we used in this calculation is that the wind is spherically symmetric [Eq.~(\ref{eq.mdot})]. Although this is likely not the case, asymmetries in the wind mass fluxes become less important at large distances to the star \citep{2014MNRAS.438.1162V}, as meridional flows aid in the latitudinal redistribution of the flow. \citet{2014MNRAS.438.1162V} also showed that winds of stars with lower surface magnetic field intensities (in fact, higher plasma-$\beta$ parameters) tend to be more homogeneous. Although the magnetic field of \gj\ has not been observationally reconstructed, its relatively low activity \citep{2015Natur.522..459E} and old age suggests lower magnetic activity \citep{2014MNRAS.441.2361V}. For the reasons stated, the mass-loss rates derived here under the assumption of spherical symmetry are believed to be robust. 

Studies that can place more firm  observationally-derived constraints on the mass-loss rates of the winds of isolated dM stars, such as our present study, are of extreme relevance to the field of winds of dM stars, as estimates of mass-loss rates of dM stars vary considerably. Here, we found that the mass-loss rate of \gj\ is $\sim [0.45,2.5] \times  10^{-15}~\msano$, which is a factor of $0.02$ to $0.13$ the present-day solar mass-loss rate ($\mdot_\odot = 2 \times 10^{-14} ~\msano$). The estimated mass-loss rates of other dM stars are shown in Table \ref{table0}.  {The mass-loss rate of \gj\ is consistent with all the upper limits presented, but it is significantly lower than those of isolated/non-interacting dM stars,  such as  EV Lac  \citep[M3.5, $1~\mdot_\odot$,][]{2005ApJ...628L.143W},  YZ CMi  \citep[M4.5, $<50~\mdot_\odot$,][]{1996ApJ...462L..91L}, and Prox Cen \citep[M5.5, $<0.2~\mdot_\odot$,][]{2001ApJ...547L..49W}, all of which are more active than \gj , as can be seen by comparing their X-ray luminosities. For interacting binary systems, values of $\mdot$ range from $0.001$ to $0.3~\mdot_\odot$ for mid- to late-dM stars \citep{2006ApJ...652..636D,2012MNRAS.420.3281P}. The mass-loss rate of \gj\ falls within this range, although it is not clear whether winds of isolated stars are similar to the winds of dM stars with very close companions. The wide range in mass-loss rates are likely due to multiple factors, such as age, magnetism (activity) and spectral types of each particular target, as well as effects of the close companion. }

Using astrospheric measurements, \citet{2002ApJ...574..412W,2005ApJ...628L.143W} showed that, for G and K dwarfs, there seems to be a relation between the mass-loss rate per unit surface area $A_\star$ and the X-ray flux $F_X$
\begin{equation}\label{eq.wood}
\mdot/A_\star \propto F_X^{1.34}, \,\,\,\, {\rm for~} F_X \lesssim 10^6~{\rm erg~cm}^{-2}{\rm s}^{-1}.
\end{equation}
If we assume that Equation (\ref{eq.wood}) can be extended to the dM regime and can be applied to \gj , which has $F_X=4.9\e4$~erg~cm$^{-2}$s$^{-1}$ (derived from an X-ray luminosity of $5.7\times10^{26}$~erg~s$^{-1}$, Table \ref{table1}), Equation (\ref{eq.wood}) would predict \gj\ to have a mass-loss rate of $\sim 4.8 \times 10^{-15}~\msano$, which is {only about a factor of 4 higher than the value derived here. Given the large uncertainties, this means that both values are roughly in line.} Note however that currently, it is not clear whether relation (\ref{eq.wood}) can be extended to dM stars, as there is no dM star with $F_X \lesssim 10^6$~erg~cm$^{-2}$s$^{-1}$ whose wind has been detected using the astrospheric technique. 

In spite of what we believe to be a robust determination of mass-loss rate with our models, our model fails to realistically 
compute the wind energetics (heating/cooling mechanisms), due to our assumption of a wind with constant temperature. Under the isothermal assumption, we derived the temperature of the wind that is required to reproduce the observed local wind velocity from \citetalias{2016A&A...591A.121B}. Our derived wind temperature ($4.1\e5$~K), however, does not agree  with the local proton temperature $\lesssim 2.4\e4$~K determined by \citetalias{2016A&A...591A.121B}. Reducing the wind temperature of our models would reduce the wind velocity at the position of the planet, no longer reproducing the observed velocities. Therefore, a simple isothermal wind model fails to reproduce simultaneously the  observed local velocity and temperature. {We also run  polytropic stellar wind models, which assume that the temperature decays with $\rho^{\gamma-1}$, where $\gamma$ is known as the polytropic index and is a free parameter. The polytropic wind model approaches the isothermal wind model as $\gamma$ approaches $1$. The polytropic index is usually assumed to lie between $1.05$ to $1.15$, as this is similar to the effective adiabatic index of $1.1$ measured in the solar wind \citep{2011ApJ...727L..32V}. In our polytropic wind models for \gj , we adopted $\gamma$ varying from $1.05$ to $1.4$ and a range of wind base temperatures. We do not show the results of these models here, as, similarly to the isothermal one, if we set the base temperature to reproduce the local wind velocity, the local temperature is larger than the measured one.} A possible way to reconcile the modelled wind temperature with the observed one would be to modify the heating/acceleration mechanism of the wind to include dissipation of MHD waves. It is expected that, after a quick increase in temperature close to the star, winds should cool down as they expand, as reproduced in MHD wave-driven wind models \citep{2006ApJ...639..416V, 2006MNRAS.368.1145F}.  Such a detailed study, which would also require  incorporating magnetic field observations of \gj , will be deferred to a future work. {Another possibility to reconcile modelled and observed temperatures would be if, somehow, the interaction between the exoplanet's exosphere would have cooled the neutralised wind material.}

The isothermal assumption also renders the inner fall-off of the wind density with distance to be less accurate. In isothermal wind models, the density decay is exponential in the subsonic region, similarly to a hydrostatic structure. Since a higher-than-observed wind temperature leads to a more extended subsonic region (up to $12$ -- $15~ R_\star$), this means that the inner density profile has a very steep dependence with distance. Put in other words, the wind base density is likely to be over-estimated in our models. The density profile is less affected in the super-sonic region and, consequently, more robust. 

Similar to mass-loss rates, the ram pressure profile beyond the planetary orbit is also robustly derived in our models. By assuming pressure balance between the ram pressures of the stellar wind and of the ISM, we estimated the distance to the termination shock in the astrosphere  of \gj\ to be $\sim 25$~au. As the sizes of astrospheres depend on the relative velocity of the stars through the ISM and on the stellar wind ram pressures, they can vary considerably from system to system. For example, the derived distances to the termination shock are around $60$~au for the dM star EV Lac, $\sim 20$~au for the K dwarf 61 Cyg A,  $\sim 800$~au for $\epsilon $ Eri (size to the termination shock in Figure 2 in \citealt{2005ApJ...628L.143W} and in Figure 2 in \citealt{2002ApJ...574..412W}) and about $80$~au for the Sun. Thus, the astrosphere of \gj\ is more compact than targets in the sample of Wood et al and the Sun.

\section{Conclusions}\label{sec.conclusions}
In this paper, we modelled the wind of the planet-hosting star \gj\ in order to derive the {\it global} properties of the stellar wind, such as the velocity and density profiles, temperature and mass-loss rates. Our wind models were constrained by a previous study of planetary evaporation \citepalias{2016A&A...591A.121B}, which derived the {\it local} density and velocity of the stellar wind (i.e., at the orbit of the transiting warm-neptune \gj b). Given its convenient location in the wind acceleration zone, \gj b was used as a probe for the wind of its host star.

Our models assumed a spherically symmetric, isothermal wind for the host star. By computing wind models with different temperatures, we were able to find the wind temperature that best reproduced the local observed wind velocity. Our results were summarised in Table~\ref{table2} and Figures \ref{fig.wind_temperature} and \ref{fig.wind_dens}. Our best-fit model has a temperature of $0.41^{+0.02}_{-0.05}$~MK, which reproduces the observed velocity of $86^{6}_{-16}$~km~s$^{-1}$ at the planet's orbit. With our models, we found that the mass-loss rate of \gj\ is $\mdot = 1.2^{+1.3}_{-0.75} \times 10^{-15} ~\msano$. Table~\ref{table0} compares the mass-loss rate we derived for \gj\ with values estimated for other dM stars, including those that are part of eclipsing binary systems. We note a  wide range of literature values of $\mdot$ derived for other dM stars ($2.2\e{-17}$ to $2\e{-14} ~\msano$), in addition to some derived upper limits. Although the mass-loss rate of \gj\ is in line with estimates of other dM stars, a more precise comparison is less suitable as differences in age, magnetism (activity) and possible effects of a close companion affect the mass-loss rate of a star.  The $\mdot$ value we derived for the M dwarf \gj\ is also well in line to the one predicted by the relation between mass-loss rates and X-ray flux \citep{2005ApJ...628L.143W} that was derived for G and K dwarfs. {As of now, the mass-loss rate of \gj\ is the weakest one ever detected for an isolated (or non-interacting) dM star.}

Because of the isothermal assumption used in our models, the inner density profile of the stellar wind (i.e., below the orbit of the planet, which roughly coincides with the sonic point) is likely to be overestimated. The density profile beyond the planet's orbit is more reliable though. Beyond the planet's orbit, we showed that the stellar wind ram pressure falls off with  $r^{-2}$. The ram pressure can be used to compute the size of planetary magnetopauses (ionopauses), in the case of magnetised planets (non-magnetised planets). It can also be used to estimate the size of the astrosphere surrounding \gj , i.e., the distance at which the wind of \gj\ reaches the interstellar medium. We found this distance to be around 25~au, which implies a compact astrosphere surrounding \gj\ (c.f., the distance to the termination shock in the heliosphere is about 80~au).

We demonstrated in this paper that transmission spectroscopy, coupled to planetary atmospheric evaporation and stellar wind models, can be a useful tool for constraining the winds of planet-hosting stars. Generalising our approach to the other planetary systems similar to \gj\ that will be discovered by upcoming transit surveys (for example, with K2, CHEOPS, TESS and PLATO missions) will open new perspectives for the combined characterisation of planetary exospheres and winds of cool dwarf stars. So far, only one planet has been confirmed around \gj . However, we reckon that transmission spectroscopy in multi-planetary systems could provide multiple constraints on velocity, density and temperature profiles of the stellar winds, if one can determine the local characteristics of the host star's wind at different distances from the star. The old K-dwarf star Kepler-444, which hosts a compact planetary system with five sub-Earth-sized exoplanets, might present such an opportunity. Kepler-444 shows strong variations in the \ly\ line that could be arising from extended hydrogen exospheres around the outer planets \citep{2017arXiv170300504B}. Modelling of the wind-exosphere interaction in these planets might present the first opportunity to derive the local stellar wind characteristics at various distances from the star. 

\section*{Acknowledgements}
VB's work has been carried out in the frame of the National Centre for Competence in Research ``PlanetS'' supported by the Swiss National Science Foundation (SNSF). VB also acknowledges the financial support of the SNSF. We thank the referee of our paper, Brian Wood, for his comments and suggestions.

\bsp
\label{lastpage}


\begin{thebibliography}{}
\makeatletter
\relax
\def\mn@urlcharsother{\let\do\@makeother \do\$\do\&\do\#\do\^\do\_\do\%\do\~}
\def\mn@doi{\begingroup\mn@urlcharsother \@ifnextchar [ {\mn@doi@}
  {\mn@doi@[]}}
\def\mn@doi@[#1]#2{\def\@tempa{#1}\ifx\@tempa\@empty \href
  {http://dx.doi.org/#2} {doi:#2}\else \href {http://dx.doi.org/#2} {#1}\fi
  \endgroup}
\def\mn@eprint#1#2{\mn@eprint@#1:#2::\@nil}
\def\mn@eprint@arXiv#1{\href {http://arxiv.org/abs/#1} {{\tt arXiv:#1}}}
\def\mn@eprint@dblp#1{\href {http://dblp.uni-trier.de/rec/bibtex/#1.xml}
  {dblp:#1}}
\def\mn@eprint@#1:#2:#3:#4\@nil{\def\@tempa {#1}\def\@tempb {#2}\def\@tempc
  {#3}\ifx \@tempc \@empty \let \@tempc \@tempb \let \@tempb \@tempa \fi \ifx
  \@tempb \@empty \def\@tempb {arXiv}\fi \@ifundefined
  {mn@eprint@\@tempb}{\@tempb:\@tempc}{\expandafter \expandafter \csname
  mn@eprint@\@tempb\endcsname \expandafter{\@tempc}}}

\bibitem[\protect\citeauthoryear{{Ben-Jaffel} \& {Sona Hosseini}}{{Ben-Jaffel}
  \& {Sona Hosseini}}{2010}]{2010ApJ...709.1284B}
{Ben-Jaffel} L.,  {Sona Hosseini} S.,  2010, \mn@doi [\apj]
  {10.1088/0004-637X/709/2/1284}, \href
  {http://adsabs.harvard.edu/abs/2010ApJ...709.1284B} {709, 1284}

\bibitem[\protect\citeauthoryear{{Bourrier} \& {Lecavelier des
  Etangs}}{{Bourrier} \& {Lecavelier des Etangs}}{2013}]{2013A&A...557A.124B}
{Bourrier} V.,  {Lecavelier des Etangs} A.,  2013, \mn@doi [\aap]
  {10.1051/0004-6361/201321551}, \href
  {http://adsabs.harvard.edu/abs/2013A%26A...557A.124B} {557, A124}

\bibitem[\protect\citeauthoryear{{Bourrier}, {Lecavelier des Etangs}  \&
  {Vidal-Madjar}}{{Bourrier} et~al.}{2014}]{2014A&A...565A.105B}
{Bourrier} V.,  {Lecavelier des Etangs} A.,   {Vidal-Madjar} A.,  2014, \mn@doi
  [\aap] {10.1051/0004-6361/201323064}, \href
  {http://adsabs.harvard.edu/abs/2014A%26A...565A.105B} {565, A105}

\bibitem[\protect\citeauthoryear{{Bourrier}, {Ehrenreich}  \& {Lecavelier des
  Etangs}}{{Bourrier} et~al.}{2015}]{2015A&A...582A..65B}
{Bourrier} V.,  {Ehrenreich} D.,   {Lecavelier des Etangs} A.,  2015, \mn@doi
  [\aap] {10.1051/0004-6361/201526894}, \href
  {http://adsabs.harvard.edu/abs/2015A%26A...582A..65B} {582, A65}

\bibitem[\protect\citeauthoryear{{Bourrier}, {Lecavelier des Etangs},
  {Ehrenreich}, {Tanaka}  \& {Vidotto}}{{Bourrier}
  et~al.}{2016}]{2016A&A...591A.121B}
{Bourrier} V.,  {Lecavelier des Etangs} A.,  {Ehrenreich} D.,  {Tanaka} Y.~A.,
   {Vidotto} A.~A.,  2016, \mn@doi [\aap] {10.1051/0004-6361/201628362}, \href
  {http://adsabs.harvard.edu/abs/2016A%26A...591A.121B} {591, A121}

\bibitem[\protect\citeauthoryear{{Bourrier} et~al.,}{{Bourrier}
  et~al.}{2017}]{2017arXiv170300504B}
{Bourrier} V.,  et~al., 2017, preprint, \href
  {http://adsabs.harvard.edu/abs/2017arXiv170300504B} {} (\mn@eprint {arXiv}
  {1703.00504})

\bibitem[\protect\citeauthoryear{{Chapman} \& {Ferraro}}{{Chapman} \&
  {Ferraro}}{1930}]{1930Natur.126..129C}
{Chapman} S.,  {Ferraro} V.~C.~A.,  1930, \mn@doi [\nat] {10.1038/126129a0},
  \href {http://adsabs.harvard.edu/abs/1930Natur.126..129C} {126, 129}

\bibitem[\protect\citeauthoryear{{Cranmer}}{{Cranmer}}{2004}]{2004AmJPh..72.1397C}
{Cranmer} S.~R.,  2004, \mn@doi [American Journal of Physics]
  {10.1119/1.1775242}, \href
  {http://adsabs.harvard.edu/abs/2004AmJPh..72.1397C} {72, 1397}

\bibitem[\protect\citeauthoryear{{Cranmer} \& {Saar}}{{Cranmer} \&
  {Saar}}{2011}]{2011ApJ...741...54C}
{Cranmer} S.~R.,  {Saar} S.~H.,  2011, \mn@doi [\apj]
  {10.1088/0004-637X/741/1/54}, \href
  {http://adsabs.harvard.edu/abs/2011ApJ...741...54C} {741, 54}

\bibitem[\protect\citeauthoryear{{Debes}}{{Debes}}{2006}]{2006ApJ...652..636D}
{Debes} J.~H.,  2006, \mn@doi [\apj] {10.1086/508132}, \href
  {http://adsabs.harvard.edu/abs/2006ApJ...652..636D} {652, 636}

\bibitem[\protect\citeauthoryear{{Dittmann} et~al.,}{{Dittmann}
  et~al.}{2017}]{2017arXiv170405556D}
{Dittmann} J.~A.,  et~al., 2017, preprint, \href
  {http://cdsads.u-strasbg.fr/abs/2017arXiv170405556D} {} (\mn@eprint {arXiv}
  {1704.05556})

\bibitem[\protect\citeauthoryear{{Ehrenreich} et~al.,}{{Ehrenreich}
  et~al.}{2012}]{2012A&A...547A..18E}
{Ehrenreich} D.,  et~al., 2012, \mn@doi [\aap] {10.1051/0004-6361/201219981},
  \href {http://adsabs.harvard.edu/abs/2012A%26A...547A..18E} {547, A18}

\bibitem[\protect\citeauthoryear{{Ehrenreich} et~al.,}{{Ehrenreich}
  et~al.}{2015}]{2015Natur.522..459E}
{Ehrenreich} D.,  et~al., 2015, \mn@doi [\nat] {10.1038/nature14501}, \href
  {http://adsabs.harvard.edu/abs/2015Natur.522..459E} {522, 459}

\bibitem[\protect\citeauthoryear{{Ekenb{\"a}ck}, {Holmstr{\"o}m}, {Wurz},
  {Grie{\ss}meier}, {Lammer}, {Selsis}  \& {Penz}}{{Ekenb{\"a}ck}
  et~al.}{2010}]{2010ApJ...709..670E}
{Ekenb{\"a}ck} A.,  {Holmstr{\"o}m} M.,  {Wurz} P.,  {Grie{\ss}meier} J.-M.,
  {Lammer} H.,  {Selsis} F.,   {Penz} T.,  2010, \mn@doi [\apj]
  {10.1088/0004-637X/709/2/670}, \href
  {http://adsabs.harvard.edu/abs/2010ApJ...709..670E} {709, 670}

\bibitem[\protect\citeauthoryear{{Falceta-Gon{\c c}alves}, {Vidotto}  \&
  {Jatenco-Pereira}}{{Falceta-Gon{\c c}alves}
  et~al.}{2006}]{2006MNRAS.368.1145F}
{Falceta-Gon{\c c}alves} D.,  {Vidotto} A.~A.,   {Jatenco-Pereira} V.,  2006,
  \mn@doi [\mnras] {10.1111/j.1365-2966.2006.10190.x}, \href
  {http://adsabs.harvard.edu/abs/2006MNRAS.368.1145F} {368, 1145}

\bibitem[\protect\citeauthoryear{{Fichtinger}, {Guedel}, {Mutel}, {Hallinan},
  {Gaidos}, {Skinner}, {Lynch}  \& {Gayley}}{{Fichtinger}
  et~al.}{2017}]{2017A&A...599A.127F}
{Fichtinger} B.,  {Guedel} M.,  {Mutel} R.~L.,  {Hallinan} G.,  {Gaidos} E.,
  {Skinner} S.~L.,  {Lynch} C.,   {Gayley} K.~G.,  2017, \mn@doi [\aap]
  {10.1051/0004-6361/201629886}, \href
  {http://cdsads.u-strasbg.fr/abs/2017A%26A...599A.127F} {599, A127}

\bibitem[\protect\citeauthoryear{{Gaidos}, {Guedel}  \& {Blake}}{{Gaidos}
  et~al.}{2000}]{2000GeoRL..27..501G}
{Gaidos} E.~J.,  {Guedel} M.,   {Blake} G.~A.,  2000, \mn@doi [\grl]
  {10.1029/1999GL010740}, \href
  {http://adsabs.harvard.edu/abs/2000GeoRL..27..501G} {27, 501}

\bibitem[\protect\citeauthoryear{{Gillon} et~al.,}{{Gillon}
  et~al.}{2017}]{2017Natur.542..456G}
{Gillon} M.,  et~al., 2017, \mn@doi [\nat] {10.1038/nature21360}, \href
  {http://cdsads.u-strasbg.fr/abs/2017Natur.542..456G} {542, 456}

\bibitem[\protect\citeauthoryear{{Grie{\ss}meier}, {Motschmann}, {Mann}  \&
  {Rucker}}{{Grie{\ss}meier} et~al.}{2005}]{2005A&A...437..717G}
{Grie{\ss}meier} J.-M.,  {Motschmann} U.,  {Mann} G.,   {Rucker} H.~O.,  2005,
  \mn@doi [\aap] {10.1051/0004-6361:20041976}, \href
  {http://adsabs.harvard.edu/abs/2005A%26A...437..717G} {437, 717}

\bibitem[\protect\citeauthoryear{{Holmstr{\"o}m}, {Ekenb{\"a}ck}, {Selsis},
  {Penz}, {Lammer}  \& {Wurz}}{{Holmstr{\"o}m}
  et~al.}{2008}]{2008Natur.451..970H}
{Holmstr{\"o}m} M.,  {Ekenb{\"a}ck} A.,  {Selsis} F.,  {Penz} T.,  {Lammer} H.,
    {Wurz} P.,  2008, \mn@doi [\nat] {10.1038/nature06600}, \href
  {http://adsabs.harvard.edu/abs/2008Natur.451..970H} {451, 970}

\bibitem[\protect\citeauthoryear{{Ip}, {Kopp}  \& {Hu}}{{Ip}
  et~al.}{2004}]{2004ApJ...602L..53I}
{Ip} W.-H.,  {Kopp} A.,   {Hu} J.-H.,  2004, \mn@doi [\apjl] {10.1086/382274},
  \href {http://adsabs.harvard.edu/abs/2004ApJ...602L..53I} {602, L53}

\bibitem[\protect\citeauthoryear{{Johnstone} \& {Guedel}}{{Johnstone} \&
  {Guedel}}{2015}]{2015A&A...578A.129J}
{Johnstone} C.~P.,  {Guedel} M.,  2015, \mn@doi [\aap]
  {10.1051/0004-6361/201425283}, \href
  {http://adsabs.harvard.edu/abs/2015A%26A...578A.129J} {578, A129}

\bibitem[\protect\citeauthoryear{{Kasting}, {Whitmire}  \&
  {Reynolds}}{{Kasting} et~al.}{1993}]{1993Icar..101..108K}
{Kasting} J.~F.,  {Whitmire} D.~P.,   {Reynolds} R.~T.,  1993, \mn@doi [Icarus]
  {10.1006/icar.1993.1010}, \href
  {http://adsabs.harvard.edu/abs/1993Icar..101..108K} {101, 108}

\bibitem[\protect\citeauthoryear{{Khodachenko} et~al.,}{{Khodachenko}
  et~al.}{2007}]{2007AsBio...7..167K}
{Khodachenko} M.~L.,  et~al., 2007, \mn@doi [Astrobiology]
  {10.1089/ast.2006.0127}, \href
  {http://adsabs.harvard.edu/abs/2007AsBio...7..167K} {7, 167}

\bibitem[\protect\citeauthoryear{{Kislyakova}, {Holmstr{\"o}m}, {Lammer},
  {Odert}  \& {Khodachenko}}{{Kislyakova} et~al.}{2014}]{2014Sci...346..981K}
{Kislyakova} K.~G.,  {Holmstr{\"o}m} M.,  {Lammer} H.,  {Odert} P.,
  {Khodachenko} M.~L.,  2014, \mn@doi [Science] {10.1126/science.1257829},
  \href {http://adsabs.harvard.edu/abs/2014Sci...346..981K} {346, 981}

\bibitem[\protect\citeauthoryear{{Knutson} et~al.,}{{Knutson}
  et~al.}{2011}]{2011ApJ...735...27K}
{Knutson} H.~A.,  et~al., 2011, \mn@doi [\apj] {10.1088/0004-637X/735/1/27},
  \href {http://adsabs.harvard.edu/abs/2011ApJ...735...27K} {735, 27}

\bibitem[\protect\citeauthoryear{{Kulow}, {France}, {Linsky}  \&
  {Loyd}}{{Kulow} et~al.}{2014}]{2014ApJ...786..132K}
{Kulow} J.~R.,  {France} K.,  {Linsky} J.,   {Loyd} R.~O.~P.,  2014, \mn@doi
  [\apj] {10.1088/0004-637X/786/2/132}, \href
  {http://adsabs.harvard.edu/abs/2014ApJ...786..132K} {786, 132}

\bibitem[\protect\citeauthoryear{{Lamers} \& {Cassinelli}}{{Lamers} \&
  {Cassinelli}}{1999}]{1999isw..book.....L}
{Lamers} H.~J.~G.~L.~M.,  {Cassinelli} J.~P.,  1999, {Introduction to Stellar
  Winds}

\bibitem[\protect\citeauthoryear{{Lammer} et~al.,}{{Lammer}
  et~al.}{2007}]{2007AsBio...7..185L}
{Lammer} H.,  et~al., 2007, \mn@doi [Astrobiology] {10.1089/ast.2006.0128},
  \href {http://adsabs.harvard.edu/abs/2007AsBio...7..185L} {7, 185}

\bibitem[\protect\citeauthoryear{{Lammer} et~al.,}{{Lammer}
  et~al.}{2009}]{2009A&ARv..17..181L}
{Lammer} H.,  et~al., 2009, \mn@doi [\aapr] {10.1007/s00159-009-0019-z}, \href
  {http://adsabs.harvard.edu/abs/2009A%26ARv..17..181L} {17, 181}

\bibitem[\protect\citeauthoryear{{Lanotte} et~al.,}{{Lanotte}
  et~al.}{2014}]{2014AA...572A..73L}
{Lanotte} A.~A.,  et~al., 2014, \mn@doi [\aap] {10.1051/0004-6361/201424373},
  \href {http://adsabs.harvard.edu/abs/2014A%26A...572A..73L} {572, A73}

\bibitem[\protect\citeauthoryear{{Lim} \& {White}}{{Lim} \&
  {White}}{1996}]{1996ApJ...462L..91L}
{Lim} J.,  {White} S.~M.,  1996, \mn@doi [\apjl] {10.1086/310038}, \href
  {http://adsabs.harvard.edu/abs/1996ApJ...462L..91L} {462, L91+}

\bibitem[\protect\citeauthoryear{{Lim}, {White}  \& {Slee}}{{Lim}
  et~al.}{1996}]{1996ApJ...460..976L}
{Lim} J.,  {White} S.~M.,   {Slee} O.~B.,  1996, \mn@doi [\apj]
  {10.1086/177025}, \href {http://adsabs.harvard.edu/abs/1996ApJ...460..976L}
  {460, 976}

\bibitem[\protect\citeauthoryear{{Llama}, {Vidotto}, {Jardine}, {Wood}, {Fares}
   \& {Gombosi}}{{Llama} et~al.}{2013}]{2013MNRAS.436.2179L}
{Llama} J.,  {Vidotto} A.~A.,  {Jardine} M.,  {Wood} K.,  {Fares} R.,
  {Gombosi} T.~I.,  2013, \mn@doi [\mnras] {10.1093/mnras/stt1725}, \href
  {http://adsabs.harvard.edu/abs/2013MNRAS.436.2179L} {436, 2179}

\bibitem[\protect\citeauthoryear{{Luhmann}}{{Luhmann}}{1986}]{1986SSRv...44..241L}
{Luhmann} J.~G.,  1986, \mn@doi [\ssr] {10.1007/BF00200818}, \href
  {http://cdsads.u-strasbg.fr/abs/1986SSRv...44..241L} {44, 241}

\bibitem[\protect\citeauthoryear{{Morin} et~al.,}{{Morin}
  et~al.}{2008}]{2008MNRAS.390..567M}
{Morin} J.,  et~al., 2008, \mn@doi [\mnras] {10.1111/j.1365-2966.2008.13809.x},
  \href {http://adsabs.harvard.edu/abs/2008MNRAS.390..567M} {390, 567}

\bibitem[\protect\citeauthoryear{{Panagia} \& {Felli}}{{Panagia} \&
  {Felli}}{1975}]{1975A&A....39....1P}
{Panagia} N.,  {Felli} M.,  1975, \aap, \href
  {http://adsabs.harvard.edu/abs/1975A%26A....39....1P} {39, 1}

\bibitem[\protect\citeauthoryear{{Parker}}{{Parker}}{1958}]{1958ApJ...128..664P}
{Parker} E.~N.,  1958, \mn@doi [\apj] {10.1086/146579}, \href
  {http://adsabs.harvard.edu/abs/1958ApJ...128..664P} {128, 664}

\bibitem[\protect\citeauthoryear{{Parsons} et~al.,}{{Parsons}
  et~al.}{2012}]{2012MNRAS.420.3281P}
{Parsons} S.~G.,  et~al., 2012, \mn@doi [\mnras]
  {10.1111/j.1365-2966.2011.20251.x}, \href
  {http://adsabs.harvard.edu/abs/2012MNRAS.420.3281P} {420, 3281}

\bibitem[\protect\citeauthoryear{{Redfield} \& {Linsky}}{{Redfield} \&
  {Linsky}}{2008}]{2008ApJ...673..283R}
{Redfield} S.,  {Linsky} J.~L.,  2008, \mn@doi [\apj] {10.1086/524002}, \href
  {http://cdsads.u-strasbg.fr/abs/2008ApJ...673..283R} {673, 283}

\bibitem[\protect\citeauthoryear{{Scalo} et~al.,}{{Scalo}
  et~al.}{2007}]{2007AsBio...7...85S}
{Scalo} J.,  et~al., 2007, \mn@doi [Astrobiology] {10.1089/ast.2006.0000},
  \href {http://adsabs.harvard.edu/abs/2007AsBio...7...85S} {7, 85}

\bibitem[\protect\citeauthoryear{{Schatzman}}{{Schatzman}}{1962}]{1962AnAp...25...18S}
{Schatzman} E.,  1962, Annales d'Astrophysique, \href
  {http://adsabs.harvard.edu/abs/1962AnAp...25...18S} {25, 18}

\bibitem[\protect\citeauthoryear{{See}, {Jardine}, {Vidotto}, {Petit},
  {Marsden}, {Jeffers}  \& {do Nascimento}}{{See}
  et~al.}{2014}]{2014A&A...570A..99S}
{See} V.,  {Jardine} M.,  {Vidotto} A.~A.,  {Petit} P.,  {Marsden} S.~C.,
  {Jeffers} S.~V.,   {do Nascimento} J.~D.,  2014, \mn@doi [\aap]
  {10.1051/0004-6361/201424323}, \href
  {http://adsabs.harvard.edu/abs/2014A%26A...570A..99S} {570, A99}

\bibitem[\protect\citeauthoryear{{Tarter} et~al.,}{{Tarter}
  et~al.}{2007}]{2007AsBio...7...30T}
{Tarter} J.~C.,  et~al., 2007, \mn@doi [Astrobiology] {10.1089/ast.2006.0124},
  \href {http://adsabs.harvard.edu/abs/2007AsBio...7...30T} {7, 30}

\bibitem[\protect\citeauthoryear{{Torres}}{{Torres}}{2007}]{2007ApJ...671L..65T}
{Torres} G.,  2007, \mn@doi [\apjl] {10.1086/524886}, \href
  {http://cdsads.u-strasbg.fr/abs/2007ApJ...671L..65T} {671, L65}

\bibitem[\protect\citeauthoryear{{Van Doorsselaere}, {Wardle}, {Del Zanna},
  {Jansari}, {Verwichte}  \& {Nakariakov}}{{Van Doorsselaere}
  et~al.}{2011}]{2011ApJ...727L..32V}
{Van Doorsselaere} T.,  {Wardle} N.,  {Del Zanna} G.,  {Jansari} K.,
  {Verwichte} E.,   {Nakariakov} V.~M.,  2011, \mn@doi [\apjl]
  {10.1088/2041-8205/727/2/L32}, \href
  {http://adsabs.harvard.edu/abs/2011ApJ...727L..32V} {727, L32}

\bibitem[\protect\citeauthoryear{{Vidotto} \& {Donati}}{{Vidotto} \&
  {Donati}}{2017}]{2017A&A...602A..39V}
{Vidotto} A.~A.,  {Donati} J.-F.,  2017, \mn@doi [\aap]
  {10.1051/0004-6361/201629700}, \href
  {http://adsabs.harvard.edu/abs/2017A%26A...602A..39V} {602, A39}

\bibitem[\protect\citeauthoryear{{Vidotto} \& {Jatenco-Pereira}}{{Vidotto} \&
  {Jatenco-Pereira}}{2006}]{2006ApJ...639..416V}
{Vidotto} A.~A.,  {Jatenco-Pereira} V.,  2006, \mn@doi [\apj] {10.1086/499329},
  \href {http://adsabs.harvard.edu/abs/2006ApJ...639..416V} {639, 416}

\bibitem[\protect\citeauthoryear{{Vidotto}, {Opher}, {Jatenco-Pereira}  \&
  {Gombosi}}{{Vidotto} et~al.}{2009}]{2009ApJ...703.1734V}
{Vidotto} A.~A.,  {Opher} M.,  {Jatenco-Pereira} V.,   {Gombosi} T.~I.,  2009,
  \mn@doi [\apj] {10.1088/0004-637X/703/2/1734}, \href
  {http://adsabs.harvard.edu/abs/2009ApJ...703.1734V} {703, 1734}

\bibitem[\protect\citeauthoryear{{Vidotto}, {Opher}, {Jatenco-Pereira}  \&
  {Gombosi}}{{Vidotto} et~al.}{2010a}]{2010ApJ...720.1262V}
{Vidotto} A.~A.,  {Opher} M.,  {Jatenco-Pereira} V.,   {Gombosi} T.~I.,  2010a,
  \mn@doi [\apj] {10.1088/0004-637X/720/2/1262}, \href
  {http://adsabs.harvard.edu/abs/2010ApJ...720.1262V} {720, 1262}

\bibitem[\protect\citeauthoryear{{Vidotto}, {Jardine}  \& {Helling}}{{Vidotto}
  et~al.}{2010b}]{2010ApJ...722L.168V}
{Vidotto} A.~A.,  {Jardine} M.,   {Helling} C.,  2010b, \mn@doi [\apjl]
  {10.1088/2041-8205/722/2/L168}, \href
  {http://adsabs.harvard.edu/abs/2010ApJ...722L.168V} {722, L168}

\bibitem[\protect\citeauthoryear{{Vidotto}, {Jardine}, {Opher}, {Donati}  \&
  {Gombosi}}{{Vidotto} et~al.}{2011}]{2011MNRAS.412..351V}
{Vidotto} A.~A.,  {Jardine} M.,  {Opher} M.,  {Donati} J.~F.,   {Gombosi}
  T.~I.,  2011, \mn@doi [\mnras] {10.1111/j.1365-2966.2010.17908.x}, \href
  {http://adsabs.harvard.edu/abs/2011MNRAS.412..351V} {412, 351}

\bibitem[\protect\citeauthoryear{{Vidotto}, {Jardine}, {Morin}, {Donati},
  {Lang}  \& {Russell}}{{Vidotto} et~al.}{2013}]{2013A&A...557A..67V}
{Vidotto} A.~A.,  {Jardine} M.,  {Morin} J.,  {Donati} J.-F.,  {Lang} P.,
  {Russell} A.~J.~B.,  2013, \mn@doi [\aap] {10.1051/0004-6361/201321504},
  \href {http://adsabs.harvard.edu/abs/2013A%26A...557A..67V} {557, A67}

\bibitem[\protect\citeauthoryear{{Vidotto}, {Jardine}, {Morin}, {Donati},
  {Opher}  \& {Gombosi}}{{Vidotto} et~al.}{2014a}]{2014MNRAS.438.1162V}
{Vidotto} A.~A.,  {Jardine} M.,  {Morin} J.,  {Donati} J.~F.,  {Opher} M.,
  {Gombosi} T.~I.,  2014a, \mn@doi [\mnras] {10.1093/mnras/stt2265}, \href
  {http://adsabs.harvard.edu/abs/2014MNRAS.438.1162V} {438, 1162}

\bibitem[\protect\citeauthoryear{{Vidotto} et~al.,}{{Vidotto}
  et~al.}{2014b}]{2014MNRAS.441.2361V}
{Vidotto} A.~A.,  et~al., 2014b, \mn@doi [\mnras] {10.1093/mnras/stu728}, \href
  {http://adsabs.harvard.edu/abs/2014MNRAS.441.2361V} {441, 2361}

\bibitem[\protect\citeauthoryear{{Villadsen}, {Hallinan}, {Bourke}, {Guedel}
  \& {Rupen}}{{Villadsen} et~al.}{2014}]{2014ApJ...788..112V}
{Villadsen} J.,  {Hallinan} G.,  {Bourke} S.,  {Guedel} M.,   {Rupen} M.,
  2014, \mn@doi [\apj] {10.1088/0004-637X/788/2/112}, \href
  {http://adsabs.harvard.edu/abs/2014ApJ...788..112V} {788, 112}

\bibitem[\protect\citeauthoryear{{Wang}, {Sheeley}, {Socker}, {Howard}  \&
  {Rich}}{{Wang} et~al.}{2000}]{2000JGR...10525133W}
{Wang} Y.-M.,  {Sheeley} N.~R.,  {Socker} D.~G.,  {Howard} R.~A.,   {Rich}
  N.~B.,  2000, \mn@doi [\jgr] {10.1029/2000JA000149}, \href
  {http://adsabs.harvard.edu/abs/2000JGR...10525133W} {105, 25133}

\bibitem[\protect\citeauthoryear{{Wargelin} \& {Drake}}{{Wargelin} \&
  {Drake}}{2002}]{2002ApJ...578..503W}
{Wargelin} B.~J.,  {Drake} J.~J.,  2002, \mn@doi [\apj] {10.1086/342270}, \href
  {http://adsabs.harvard.edu/abs/2002ApJ...578..503W} {578, 503}

\bibitem[\protect\citeauthoryear{{Weber} \& {Davis}}{{Weber} \&
  {Davis}}{1967}]{1967ApJ...148..217W}
{Weber} E.~J.,  {Davis} L.~J.,  1967, \mn@doi [\apj] {10.1086/149138}, \href
  {http://adsabs.harvard.edu/abs/1967ApJ...148..217W} {148, 217}

\bibitem[\protect\citeauthoryear{{Wood}}{{Wood}}{2004}]{2004LRSP....1....2W}
{Wood} B.~E.,  2004, Living Reviews in Solar Physics, \href
  {http://adsabs.harvard.edu/abs/2004LRSP....1....2W} {1, 2}

\bibitem[\protect\citeauthoryear{{Wood} \& {Linsky}}{{Wood} \&
  {Linsky}}{1998}]{1998ApJ...492..788W}
{Wood} B.~E.,  {Linsky} J.~L.,  1998, \mn@doi [\apj] {10.1086/305077}, \href
  {http://adsabs.harvard.edu/abs/1998ApJ...492..788W} {492, 788}

\bibitem[\protect\citeauthoryear{{Wood}, {Linsky}, {M{\"u}ller}  \&
  {Zank}}{{Wood} et~al.}{2001}]{2001ApJ...547L..49W}
{Wood} B.~E.,  {Linsky} J.~L.,  {M{\"u}ller} H.,   {Zank} G.~P.,  2001, \mn@doi
  [\apjl] {10.1086/318888}, \href
  {http://adsabs.harvard.edu/abs/2001ApJ...547L..49W} {547, L49}

\bibitem[\protect\citeauthoryear{{Wood}, {M{\"u}ller}, {Zank}  \&
  {Linsky}}{{Wood} et~al.}{2002}]{2002ApJ...574..412W}
{Wood} B.~E.,  {M{\"u}ller} H.-R.,  {Zank} G.~P.,   {Linsky} J.~L.,  2002,
  \mn@doi [\apj] {10.1086/340797}, \href
  {http://adsabs.harvard.edu/abs/2002ApJ...574..412W} {574, 412}

\bibitem[\protect\citeauthoryear{{Wood}, {M{\"u}ller}, {Zank}, {Linsky}  \&
  {Redfield}}{{Wood} et~al.}{2005}]{2005ApJ...628L.143W}
{Wood} B.~E.,  {M{\"u}ller} H.-R.,  {Zank} G.~P.,  {Linsky} J.~L.,   {Redfield}
  S.,  2005, \mn@doi [\apjl] {10.1086/432716}, \href
  {http://adsabs.harvard.edu/abs/2005ApJ...628L.143W} {628, L143}

\bibitem[\protect\citeauthoryear{{Zendejas}, {Segura}  \& {Raga}}{{Zendejas}
  et~al.}{2010}]{2010Icar..210..539Z}
{Zendejas} J.,  {Segura} A.,   {Raga} A.~C.,  2010, \mn@doi [\icarus]
  {10.1016/j.icarus.2010.07.013}, \href
  {http://adsabs.harvard.edu/abs/2010Icar..210..539Z} {210, 539}

\bibitem[\protect\citeauthoryear{{do Nascimento} Jr. et~al.,}{{do Nascimento}
  et~al.}{2016}]{2016ApJ...820L..15D}
{do Nascimento} Jr. J.-D.,  et~al., 2016, \mn@doi [\apjl]
  {10.3847/2041-8205/820/1/L15}, \href
  {http://adsabs.harvard.edu/abs/2016ApJ...820L..15D} {820, L15}

\bibitem[\protect\citeauthoryear{{van Leeuwen}}{{van
  Leeuwen}}{2007}]{2007A&A...474..653V}
{van Leeuwen} F.,  2007, \mn@doi [\aap] {10.1051/0004-6361:20078357}, \href
  {http://adsabs.harvard.edu/abs/2007A%26A...474..653V} {474, 653}

\bibitem[\protect\citeauthoryear{{van den Oord} \& {Doyle}}{{van den Oord} \&
  {Doyle}}{1997}]{1997AandA...319..578V}
{van den Oord} G.~H.~J.,  {Doyle} J.~G.,  1997, \aap, \href
  {http://adsabs.harvard.edu/abs/1997A%26A...319..578V} {319, 578}

\makeatother
\end{thebibliography}
\end{document}